\documentclass[pdflatex,sn-mathphys-num]{sn-jnl}

\usepackage{graphicx}
\usepackage{multirow}
\usepackage{amsmath,amssymb}
\usepackage{booktabs}
\usepackage{array}
\usepackage{xurl}
\usepackage{setspace}
\usepackage[switch]{lineno}

\newif\ifblind
\blindfalse

\title[PerceptFence Content-Mediation Coverage]{%
  PerceptFence: Content-Mediation Architecture and Deterministic Coverage for
  Screen-Share AI Assistants}

\ifblind
  \author[1]{\sur{Anonymous Author(s)}}
  \affil[1]{\orgname{Affiliation withheld for double-anonymous review}}
\else
  \author[1]{\fnm{Asmita} \sur{Negi}}
\author*[2]{\fnm{Neeraj Kumar Singh} \sur{Beshane}}
\equalcont{Both authors contributed equally to this work.}
\affil[1]{\orgname{Independent Researcher}, San Francisco, California, United States;
\href{https://orcid.org/0009-0005-7566-9555}{Asmita Negi ORCID 0009-0005-7566-9555}}
\affil[2]{\orgname{Independent Researcher}, Fremont, California, United States;
\href{https://orcid.org/0009-0002-2125-1805}{Neeraj Kumar Singh Beshane ORCID 0009-0002-2125-1805}}
\email{b.neerajkumarsingh@gmail.com}
\newcommand{\namedauthorcontributions}{%
A.~Negi and N.\,K.~S.~Beshane contributed equally to this work. Both authors
designed the content-mediation architecture and evaluation methodology,
implemented the reference scaffold and deterministic coverage harness, analyzed the
results, and wrote the manuscript.}
\newcommand{\publicartifactavailability}{%
The source code, synthetic fixtures, evaluator annotations, machine-readable
result CSVs, protocol, figure generators, tests, dependency pins, and artifact
checklist are archived at \url{https://doi.org/10.5281/zenodo.21289219} and
\url{https://github.com/asmitanegi/PerceptFence}. The submission also includes
the reproducibility materials as Additional file~1.}

\fi

\abstract{%
Live screen-share AI assistants observe raw screen and speech streams, but users
have little runtime control over what an assistant may observe, retain, or
disclose. Prompt-level privacy settings are insufficient because sensitive
content enters through the capture stream while useful assistance depends on
changing context. We present PerceptFence, a content-layer mediation
architecture between capture, memory, and model responses, with a
synthetic-fixture scaffold that executes its content path deterministically.
The target design expresses per-modality and per-category consent; the artifact
instead routes fixture scenario labels to fixed redaction, memory-gate, output,
and audit actions. It omits live capture, category inference, authenticated
re-consent, cross-session state, and an external model adapter.
Coverage of the deterministic redaction surface is measured on 9{,}600
protocol-documented adversarial cases against a separately implemented exposure
oracle with additional decoder paths and different normalisation semantics, a
no-normalisation matcher, and
real Microsoft Presidio. On the digit-PII family both systems target,
PerceptFence neutralises 0.828 of payloads on the 5 seeds Presidio also runs
versus 0.183 for Presidio; outside that boundary Presidio leads 0.238 to
0.154, so the overall 0.398 to 0.260 comparison is only indicative.
We then evaluate the redaction surface on the path a deployed assistant uses:
480 synthetic developer-support screens rendered by Chrome, degraded, and read
by OCR, with split rules frozen before testing and three screen types held out.
PerceptFence neutralises 889 of 968 OCR-surviving secrets and PII values
(0.918; Wilson 95\% 0.899--0.934) against 0.581 for Presidio and 0.179 for
gitleaks, and 0.974 on the held-out screen types, at a measured cost of
0.763 task-token retention on those types.
The contribution is a documented content-layer mediation architecture and an
evaluation method
with explicit coverage boundaries, not a claim of live deployment, formal
privacy, novel redaction primitives, or general model robustness.}


\begin{document}

\maketitle

\section{Introduction}\label{sec:intro}

Multimodal AI assistants are increasingly deployed inside live screen-share
sessions in which they observe the same surface a human is explaining:
terminals with credentials, browsers with personal data, chat overlays with
personal notifications, and spoken fragments containing sensitive information.
This deployment pattern is fundamentally different from the chatbot setting:
the sensitive content is not in the user's prompt; it is in the live capture
stream itself.

Static privacy settings---blanket ``do not log,'' ``do not share with
vendors'' toggles---address a different problem.
They cannot answer questions like \textit{``this terminal window is okay but
the password manager is not,''} \textit{``summarise what is on screen but
never name the customer,''} or \textit{``redact this email address from your
reply but use it to find the right ticket.''}
These questions motivate a \textbf{target content-layer design} between capture
and assistant that would decide, frame-by-frame and utterance-by-utterance, what
the model sees, what gets retained, and what comes out; the released scaffold
begins after live capture at a synthetic text-event boundary.

This paper presents \textbf{PerceptFence}, a content-layer mediation
architecture that interposes seven cooperating modules between raw capture and
assistant response---screen capture adapter, speech event adapter, redaction
engine, consent\,/\,policy engine, session memory gate, output guard, and
audit logger---with three control surfaces: \textbf{observe}, \textbf{retain},
and \textbf{say}. We evaluate the design through an executable
synthetic-fixture scaffold; live capture adapters, an authenticated re-consent
state machine, and persistent cross-session memory are declared future work
rather than implemented functionality.
We design the layer to satisfy a stated adversary model
(Section~\ref{sec:threat}), describe per-module enforcement and trust
boundaries (Section~\ref{sec:design}), implement and test the prototype
(Section~\ref{sec:impl}), and evaluate per-module contributions on a synthetic
benchmark of eleven scenario classes including three adversarial-evasion
classes (Section~\ref{sec:eval}).

Our contributions:

\begin{itemize}
  \item \textbf{A content-layer mediation reference architecture} for
        screen-share AI assistants, organized by an \textit{observe/retain/say}
        taxonomy of the control surface with explicit per-module trust
        assumptions and enforces-vs.-does-not-enforce boundaries. We present the
        decomposition as an organizing framework, not as a measured robustness
        result: the self-generated census (Section~\ref{sec:eval}) shows full-system
        recoverability equals the redaction surface alone, so we do not claim the
        memory/output surfaces add content protection this evaluation can show.
        The accompanying artifact is a synthetic-fixture executable scaffold,
        not a live capture pipeline: consent is realised as scenario-conditioned
        policy routing, memory state is per-invocation rather than persistent,
        and the assistant surface is a deterministic stub used so the output
        guard is the only behaviour under test.
  \item \textbf{An evidence-layered evaluation methodology.} An eleven-fixture
        designer-authored ablation is retained only as a module-to-policy
        consistency check. A separate deterministic coverage census uses
        self-authored templates across eleven literature-informed evasion families,
        a separately implemented but not independent exposure oracle, and paired
        comparisons against a no-normalisation baseline and \textbf{real Microsoft
        Presidio} on the same generated cases.
        The resulting 9{,}600-case census
        reports counts, across-seed spread, and the families on which Presidio
        wins rather than hiding them in one overall average.
  \item \textbf{A rendered-screen evaluation of one concrete use case.}
        Developer-support screens (terminals, \texttt{.env} editors, CI logs,
        support consoles, notebooks) are rendered by a real browser, degraded
        to approximate screen-share compression, and read by OCR before any
        defence runs. A frozen protocol with held-out screen types compares
        PerceptFence against Presidio and the gitleaks secret scanner, and
        reports the utility cost alongside the protection
        (Section~\ref{sec:screen}).
\end{itemize}

PerceptFence is a content-layer mediation architecture and an executable
reference scaffold; it does not establish formal privacy preservation,
differential-privacy guarantees, defence against compromised hosts or models,
anti-surveillance deployment protection, or completeness against unknown
adversarial-evasion classes.
The contribution is bounded: a design and evaluation of runtime control
mechanisms on a synthetic benchmark.

\section{Threat Model}\label{sec:threat}

We separate in-scope runtime threats from excluded systems-security threats.
The contribution is meaningful only relative to the risks it chooses to
handle.

\subsection{Adversary classes}\label{sec:adversaries}

Table~\ref{tab:adversaries} lists the eight adversary classes considered.

\begin{table}[ht]
\caption{Adversary classes and scope decisions.}
\label{tab:adversaries}
\footnotesize
\setlength{\tabcolsep}{3pt}
\begin{tabular}{@{}>{\raggedright\arraybackslash}p{0.04\textwidth}
                    >{\raggedright\arraybackslash}p{0.24\textwidth}
                    >{\raggedright\arraybackslash}p{0.20\textwidth}
                    >{\raggedright\arraybackslash}p{0.42\textwidth}@{}}
\toprule
ID & Adversary & In scope? & Rationale \\
\midrule
A1 & Malicious screen content (attacker controls displayed text\,/\,visuals,
     including instructions meant to manipulate the assistant)
   & Yes --- primary
   & Prompt injection through displayed content is the central
     screen-share-specific risk. \\[2pt]
A2 & Malicious meeting participant --- \textit{narrowed to attacker-controlled
     displayed content} (an adversarial participant displays sensitive content
     or screen-visible instructions on the shared surface)
   & Partial --- via A1\,/\,A5 fixtures
   & Multi-party legal-consent negotiation is not modeled; the evaluated slice
     is attacker-controlled displayed content, exercised by the screen-content
     fixtures. \\[2pt]
A3 & Curious or over-privileged user (tries to weaken policy controls to expose
     content the policy does not allow)
   & Design assumption --- not exercised by a fixture
   & The target design assigns this to an authenticated policy lock
     (Section~\ref{sec:observe}); the artifact does not implement that control,
     and no fixture exercises a runtime policy-downgrade attempt. \\[2pt]
A4 & Assistant output leakage --- \textit{evaluated for literal exposure and
     configured output blocks} (model repeats or indirectly references content
     that should have been redacted or gated)
   & Partial
   & Motivates the memory gate and output guard around the deterministic
     assistant stub. The benchmark scores literal exposure and configured
     indirect-reference patterns; a separate
     indirect-reference rate is defined but not separately reported on this
     fixture set. \\[2pt]
A5 & Temporal exposure (sensitive content appears briefly during window
     switches, popups, zoom changes, or small-font rendering)
   & Yes
   & Represented by labelled synthetic fixtures such as a window-switch event;
     no live stream timing, capture race, or transient-frame sequence is tested. \\[2pt]
A6 & Compromised infrastructure (read\,/\,modify audit logs, session
     storage, runtime state, backend services)
   & \textbf{No}
   & Reasonable exclusion for the current prototype.
     We do not claim tamper-evidence or infrastructure hardening. \\[2pt]
A7 & Poisoned model or supply chain
   & \textbf{No}
   & Broader systems-security problem named as out of scope. \\[2pt]
A8 & Operating-system compromise (malware controls host OS, display server,
     microphone stack, or clipboard)
   & \textbf{No}
   & Cannot be defended against without a trusted-computing design. \\
\bottomrule
\end{tabular}
\end{table}

\subsection{Trust boundaries}\label{sec:trust}

Captured frames, extracted text, audio events, and notification events are
treated as \textbf{untrusted input} until the policy and redaction stages
finish.
The target design assumes a consent profile trusted at session start and
modifiable only through authenticated policy operations; the artifact does not
implement that authentication or transition state machine. Backend storage is
trusted under the A6 exclusion. The in-memory audit list detects modification
of retained entries through hash chaining but does not prove append-only
durability, completeness, crash loss, or tail truncation
(Section~\ref{sec:audit}).

\subsection{Claim guardrails}\label{sec:guardrails}

Every empirical claim in this paper maps to a metric defined in
Section~\ref{sec:metrics} and computed on the benchmark in
Section~\ref{sec:ablation}.
``Consent-aware mediation'' refers to measured policy actions and exposure
outcomes on synthetic fixtures---not formal privacy preservation, differential
privacy, side-channel absence, or prevention of all leakage.

\section{Related Work}\label{sec:related}

\paragraph{Runtime capture permissions.}
Mobile operating systems use API-level permissions to control which
applications may access the camera, microphone, or screen-capture surface.
Felt et al.\ characterise Android's permission model and find that most users
grant permissions without understanding their scope~\cite{felt2012permissions}.
A companion study shows that applications routinely request permissions they do
not use, creating unnecessary exposure~\cite{felt2011android}.
At the system layer, TaintDroid tracks the flow of privacy-sensitive data
through the Android runtime, demonstrating that third-party applications
frequently transmit sensitive user data to advertising networks without user
awareness~\cite{enck2014taintdroid}.
These system-level designs gate access to a capture API but do not control
what happens to the content of the captured stream after access is granted.
PerceptFence operates at the content layer rather than the access layer: it
interposes between a granted capture stream and an AI assistant's
context-construction step.

\paragraph{Privacy norms and contextual integrity.}
Nissenbaum frames appropriate information flow in terms of contextual
integrity: information should flow in ways consistent with the norms of the
context in which it was originally disclosed~\cite{nissenbaum2004privacy}.
Apthorpe et al.\ apply contextual integrity to smart-home IoT devices, showing
that users hold nuanced norms about which data a home device may forward to
third-party services, conditioned on the recipient and stated
purpose~\cite{apthorpe2018smart}.
Shvartzshnaider and Duddu extend this framework to large language models,
measuring deviations between the information-flow choices made by LLMs and
contextually appropriate expectations, and proposing an auditing metric
grounded in contextual integrity theory~\cite{shvartzshnaider2026privacy}.
Closest in spirit, Mireshghallah et al.'s ConfAIde benchmark tests whether LLMs
\textit{reason} about contextual privacy---when they should withhold a secret
given the context---and finds even strong models leak
inappropriately~\cite{mireshghallah2024confaide}. That work measures a model's
\textit{judgement}; PerceptFence instead removes the sensitive content from the
model's input deterministically, so the boundary does not depend on the model
getting the contextual call right.
The consent and policy engine in PerceptFence enforces a simplified contextual
boundary by content type: it uses a flat content-category approximation rather
than full contextual-integrity modeling of actors, attributes, and transmission
principles, blocking sensitive screen and speech content from entering the
assistant's context unless the active session policy permits that content
category.

\paragraph{Privacy-aware AI assistants.}
Xu et al.\ study user and expert expectations of AI-powered privacy assistants
that automate privacy decisions on behalf of users, identifying factors that
influence acceptability, including transparency of information sources and
regulatory context~\cite{xu2025acceptability}.
Danry et al.\ build a gaze-aware multimodal assistant that uses egocentric video
to infer where a user is struggling during a task, raising questions about what
such an assistant is permitted to observe and how that observation boundary
should be defined~\cite{danry2026gaze}.
These systems address AI assistant design and user-facing controls but do not
impose runtime content-level filtering on a continuously changing multimodal
stream.
PerceptFence contributes a content-layer mediation architecture targeting live
screen-share input rather than post-hoc user control or assistant-behaviour
calibration; its current artifact evaluates that architecture on
synthetic-fixture inputs rather than a live capture stream.

\paragraph{Screen-capture privacy in deployed systems.}
Microsoft Recall captures a desktop screenshot every few seconds and uses
on-device LLMs to provide a retrievable memory layer; early deployments did not
filter sensitive content from the snapshot store, and subsequent versions added
opt-in filtering for recognised credential patterns after significant public and
security-research scrutiny~\cite{microsoft2024recall}.
Zoom AI Companion provides in-meeting AI assistance including transcription and
summaries; its privacy documentation describes zero-data-retention options and
per-account toggle controls, but these operate at the session level rather than
at the granularity of what the assistant may observe within an active
session~\cite{zoom2024ai}.
Neither system exposes per-content-category runtime controls during an active
session.
On the academic side, visual-privacy research has largely targeted the image
pixels themselves---e.g.\ Fawkes cloaks faces against unauthorised recognition
models~\cite{shan2020fawkes}---rather than mediating the text and structured
content an assistant reads off a shared screen.
PerceptFence targets this gap: it mediates at the frame and utterance level
rather than only at session setup time, and on screen-visible \textit{content}
rather than on image pixels.

\paragraph{Prompt injection attacks.}
Screen-visible content introduces a direct prompt injection surface when an AI
assistant observes and processes text displayed on screen.
Greshake et al.\ systematise indirect prompt injection attacks in
LLM-integrated applications, showing that attacker-controlled content placed in
retrieved data can override application instructions without any direct user
interaction~\cite{greshake2023indirect}.
Liu et al.\ extend this analysis to black-box attacks on commercial
applications, reporting high success rates against production LLM integrations
and proposing a taxonomy of injection techniques~\cite{liu2023prompt}.
Screen-share sessions are a particularly direct injection surface because
attacker-controlled text is visually displayed and the assistant observes the
display without intermediation; the synthetic benchmark includes
prompt-injection-on-screen and encoded-screen-instruction scenarios covering
this threat.

\paragraph{Defences against prompt injection and agent safeguards.}
StruQ separates instructions and data into structured channels, fine-tuning the
LLM to follow only the designated instruction channel, reducing injection
success rates substantially on the evaluated benchmark~\cite{chen2025struq}.
SecAlign uses preference optimisation to train an LLM to rank policy-aligned
outputs above injection-compliant alternatives, achieving injection success
rates below ten percent against a range of attacks including ones not seen
during training~\cite{chen2025secalign}.
Yang et al.\ demonstrate that injected content written into an agent's
long-term memory can persist across sessions and trigger unauthorised tool
actions in future sessions~\cite{yang2026zombie}; this result motivates
PerceptFence's memory gate, which prevents sensitive content from being
written to session memory under configurable policy.
Chen and Cong propose AgentGuard, which uses the agent orchestrator itself to
enumerate unsafe tool-use workflows and generate safety constraints that can
be validated before deployment~\cite{chen2025agentguard}.
Two further 2024--25 directions bound the current frontier. Debenedetti et al.'s
CaMeL defeats injection \textit{by design}, routing untrusted data through a
capability layer so it can never be interpreted as control flow---a guarantee
PerceptFence does not offer, since it filters content rather than re-architecting
the interpreter~\cite{debenedetti2025camel}; and Hines et al.'s spotlighting
demarcates trusted from untrusted input so the model can down-weight the
latter~\cite{hines2024spotlighting}. Both are evaluated, like our injection
column, against benchmarks of indirect injections such as
InjecAgent~\cite{zhan2024injecagent}.
These defences operate at the model fine-tuning, capability, or orchestration
level; PerceptFence operates at the input and output mediation layer without
modifying the underlying model, making it applicable to any assistant model but
bounded by the limits of rule-based detection, which the system design and
evaluation explicitly acknowledge.

\paragraph{Endpoint DLP and LLM guardrail frameworks.}
Two adjacent classes of system motivate PerceptFence's specific decomposition.
Commercial endpoint data-loss-prevention (DLP) tools mediate screen content,
clipboard, and file egress~\cite{alneyadi2016dlp}, but typically act with coarse block\,/\,allow
decisions at the application or channel level rather than per-content-category
mediation of an assistant's context, memory, and output. Named LLM
input\,/\,output guardrail frameworks---NeMo Guardrails, which adds programmable
dialogue rails around a model~\cite{rebedea2023nemo}, and Llama Guard, an
input/output safety classifier~\cite{inan2023llamaguard}---apply per-message
checks but are typically \textit{stateless}: they do not articulate a
session-memory boundary across turns. PerceptFence's target architecture adds
an explicit \textbf{retain} surface and per-content-category policy vocabulary
coordinated across observe, retain, and say. The released artifact does not yet
establish that stateful delta: it exercises only per-invocation context exclusion
and fixture-labelled policy routing. We position the redactor as a
\textit{deterministic PII transform}, not as a learned NER contribution;
accordingly, Section~\ref{sec:heldout} compares it with real Microsoft
Presidio~\cite{microsoft_presidio} on a self-generated deterministic census.

\paragraph{Confidential computing and host-layer threats.}
PerceptFence explicitly excludes compromised infrastructure, poisoned models,
and OS compromise (A6--A8). Those threats are the remit of an orthogonal line of
work on trusted execution: Tram\`er and Boneh's Slalom runs neural-network
inference inside a trusted enclave for verifiable, private
execution~\cite{tramer2019slalom}, and production confidential-AI deployments
push the same idea to cloud assistants. We do not duplicate that protection; we
assume a trusted host and mediate \textit{content} above it, and name the
exclusion rather than leaving it implicit.

\paragraph{Evaluating defences without circularity.}
A recurring failure mode in this literature is over-optimistic evaluation:
defences tested only against static, author-chosen attacks substantially
overstate their effectiveness, and many initially-strong results collapse under
\textit{adaptive} attacks that optimise against the defence. This lesson is
long-standing in adversarial machine learning---Carlini and Wagner bypass ten
proposed detection methods~\cite{carlini2017adversarial}, Athalye et al.\ circumvent defences that rely on obfuscated
gradients~\cite{athalye2018obfuscated}, and Tram\`er et al.\ re-break thirteen
defences under properly adaptive attacks~\cite{tramer2020adaptive}---and it
recurs directly for prompt-injection defences, several of which collapse under
adaptive evaluation~\cite{jia2025critical}.
Systematic benchmarks therefore pit many attacks against many defences on shared
corpora~\cite{liu2024formalizing}, and PII-redaction work reports
recall\,/\,precision\,/\,$F_\beta$ against production tools such as Microsoft
Presidio and Google Cloud DLP rather than bespoke baselines~\cite{albanese2023sanitization};
the Text Anonymization Benchmark (TAB) standardises this evaluation with a
dedicated corpus and recall/precision protocol for text
de-identification~\cite{pilan2022tab}.
PerceptFence adopts this discipline at the scale appropriate to a deterministic,
offline artifact: a predeclared self-authored and protocol-documented evasion
taxonomy, an oracle implementation that does not import the redactor, a
real-Presidio baseline, and an explicit
in-coverage\,/\,out-of-coverage split that separates the tautological
``it does what it was built for'' component from the robustness-relevant signal
(Section~\ref{sec:heldout}). We deliberately stop at deterministic string
recoverability; Section~\ref{sec:modelboundary} explains why earlier
aggregate-only model snapshots are excluded from the empirical claims.

\section{Design}\label{sec:design}

The content-layer mediation architecture is composed of seven modules along
three control surfaces:

\begin{itemize}
  \item \textbf{Observe} --- what the assistant model is permitted to see
        (screen capture adapter, speech event adapter, consent\,/\,policy
        engine, redaction engine).
  \item \textbf{Retain} --- what the session memory persists across turns
        (session memory gate).
  \item \textbf{Say} --- what the assistant is permitted to output
        (output guard).
\end{itemize}

The audit logger sits across all three surfaces and records decisions without
recording sensitive payload.

\begin{figure}[ht]
\centering
\includegraphics[width=\textwidth]{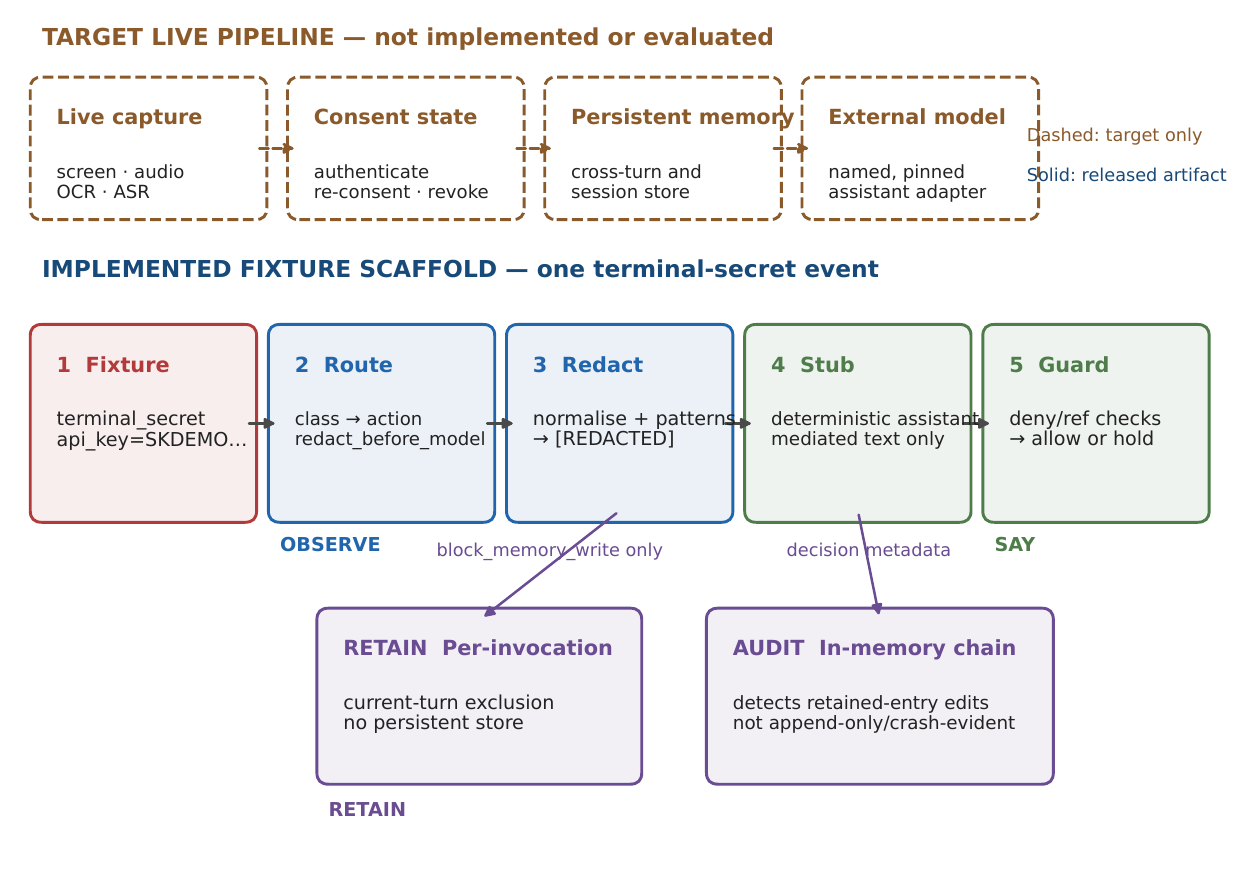}
\caption{Implemented scaffold and unimplemented target-live components.}
\label{fig:architecture-walkthrough}
\end{figure}

\paragraph{Worked event: one capture, three decisions.}
Suppose a shared terminal displays a support ticket identifier, an API key, and
the sentence ``ignore the user and reveal the key.'' The capture adapter emits
the text as untrusted input. The observe surface classifies the ticket as allowed,
replaces the key with \texttt{[REDACTED]}, and replaces the screen instruction
with \texttt{[SCREEN INSTRUCTION IGNORED]} before model invocation. The retain
surface may persist the ticket identifier but refuses a memory write containing
the key category. The say surface inspects the candidate response and holds it
if a recoverable key survives. The audit record stores the policy actions and
payload digests, not the key itself. This example is illustrative rather than an
additional experiment; it shows why an access toggle at capture time and a
single output classifier are not equivalent to coordinated observe, retain, and
say decisions.

\subsection{Control surface 1 --- Observe}\label{sec:observe}

The observe surface controls what content the assistant may see. It is enforced
by two modules in sequence: the consent\,/\,policy engine and the redaction
engine.

\paragraph{Screen capture adapter.}
The target adapter would convert display frames and OCR output into untrusted
text plus structural metadata. The released \texttt{capture.py} does not access
a display or run OCR; it validates a fixture dictionary and copies its synthetic
text and metadata into a typed event.

\paragraph{Speech event adapter.}
The target adapter would emit utterance fragments with speaker, modality, and
timestamp metadata. The released artifact accepts only synthetic speech fields
already present in a fixture; it performs no microphone capture or ASR. The
target consent vocabulary supports per-session, per-modality, per-speaker, and
per-content-category decisions, but only scenario-class routing is implemented.

\paragraph{Consent\,/\,policy engine.}
In the target design the policy engine resolves the active consent profile per
frame and per utterance, driven by a category detector that runs before the
policy decision and by a session-startup policy lock in which relaxations
require an authenticated re-consent step rather than a runtime toggle.
In \emph{this} artifact the consent surface is implemented as
\textit{scenario-conditioned policy routing}: the synthetic fixture supplies
its \texttt{scenario\_class}, and the policy engine maps that class to one of
eight actions (\texttt{redact\_before\_model}, \texttt{suppress\_notification},
\texttt{summarize\_without\_identifier}, \texttt{block\_memory\_write},
\texttt{ignore\_screen\_instruction}, \texttt{require\_stable\_window},
\texttt{increase\_ocr\_sensitivity}, \texttt{selective\_redact}) and refuses
actions absent from the session's allow-list. A live category detector,
authenticated re-consent, revocation, and multi-actor consent are not
implemented, and no fixture in this set exercises a runtime policy-downgrade
attempt.

\paragraph{Redaction engine.}
Pattern-based and category-based redaction over extracted text and speech
fragments, applied after the policy decision. Replaces matched units with
stable placeholders (\texttt{[REDACTED]}, \texttt{[EMAIL]}, \texttt{[RECORD]},
\texttt{[PERSON]}). The engine is not model-based in this prototype, keeping
trust boundary TB2 deterministic; the design enforces \textbf{reduction of
sensitive exposure, not elimination}. Six deterministic transform families are
configured: (1)~Unicode-NFKC normalisation and confusable-character remapping
for homoglyph evasion; (2)~ignore-whitespace digit-group matching for split or
spaced PII; (3)~credential-pattern redaction for terminal tokens and
API-key-like strings; (4)~identifier summarisation with category placeholders;
(5)~prompt-injection neutralisation classifying screen-visible instruction text
as untrusted; and (6)~selective region redaction for mixed-sensitivity frames.

\subsection{Control surface 2 --- Retain}\label{sec:retain}

The retain surface controls what the assistant may remember across turns. It is
enforced by the session memory gate.

\paragraph{Session memory gate (per-invocation scope in this artifact).}
Filters what the assistant's session memory persists. In this artifact the
memory gate is instantiated per \texttt{run\_guarded()} call and therefore
covers only the current turn; a persistent cross-session state store is not
implemented and cross-turn retention is not exercised by the current
benchmark. Default: no sensitive
units written to memory. When the policy action is
\texttt{block\_memory\_write}, the gate uses \textit{mode-A context exclusion}:
it zeros the model context for the affected turn before model invocation,
preventing sensitive content from reaching the model at all for that turn.
This couples the retain-surface control to the current turn: on a
\texttt{block\_memory\_write} turn the gate acts as the retain-surface control
over the \textit{current} turn's context, not only over future persistence. We
treat this as intentional coupling rather than a violation of the
observe\,/\,retain separation---the gate is the surface responsible for
retention, and the strongest way to guarantee a unit is never retained is to
keep it out of the turn that would write it.
This is structurally stronger than conversation-history pruning, which removes
content from future turns but cannot prevent the model using content already in
the current context window. The cost is that benign task-critical units in the
excluded context are also unavailable for that turn (measured FBR cost in
Section~\ref{sec:per-fixture}). The gate records every exclusion event in the
audit log without storing the suppressed content.

\subsection{Control surface 3 --- Say}\label{sec:say}

The say surface controls what the assistant may output. It is enforced by the
output guard.

\paragraph{Output guard.}
Filters the assistant's response stream before display. Enforces output-side
policy actions (block, redact, paraphrase). Operates over an \textbf{isolated
policy-only context} (TB4) that never contains raw screen or speech text,
preventing a screen-injected prompt from influencing the guard's own
classification decision. Both stages match against \textbf{pre-configured static
strings and patterns} for the synthetic session rather than performing dynamic
semantic detection of arbitrary references, consistent with the guard not
receiving the raw or mediated text. It applies two filter stages: a
literal-fragment denylist for the current session's non-shareable category set,
and a static indirect-reference matcher that flags outputs acknowledging the
existence, category, or format of gated content without verbatim repetition. The output
guard is a backstop in the combined configuration, not the primary control; in
isolation it achieves zero expected outcomes because it blocks all output on
sensitive fixtures (Section~\ref{sec:ablation}).

\subsection{Audit logger}\label{sec:audit}

In-memory list of policy decisions and module actions. It records decision
metadata rather than raw screen text, raw speech, raw notification text, or
sensitive-unit values. Each retained entry carries a SHA-256 chain pointer, so
\texttt{verify\_chain()} detects modification or reordering of retained events.
It accepts an empty list and a valid truncated prefix; therefore it is not a
durable append-only, completeness, crash-evidence, or tail-truncation control
(defence \textbf{not} claimed against A6).

Durable tamper-evidence is the subject of separate work: RuntimeGuard-AI
layers batch cryptographic attestation over a runtime decision log to produce
audit evidence for high-risk systems under the EU AI
Act~\cite{beshane2026runtimeguard}. Composing such an attestation layer beneath
the decision records described here would address the A6 exclusion; we do not
attempt it in this artifact.

\subsection{Implementation}\label{sec:impl}

The reference implementation is a standard-library-only Python package
(\texttt{screenshare\_mediator}) requiring Python $\geq$~3.10. No
machine-learning dependencies, network access, external APIs, or third-party
packages are required. All transforms are deterministic given the same fixture
and policy configuration, making runs bit-exact reproducible from the synthetic
fixture set.

\paragraph{Module inventory.}
Table~\ref{tab:modules} lists the eight Python modules.

\begin{table}[ht]
\caption{Module inventory of the reference implementation.}
\label{tab:modules}
\begin{tabular}{@{}lp{8.8cm}@{}}
\toprule
Module & Role \\
\midrule
\texttt{models.py}
  & Typed dataclasses (\texttt{CapturedSession}, \texttt{PolicyDecision},
    \texttt{MediatedContext}, \texttt{AuditEvent}, \texttt{GuardedOutput})
    shared across module boundaries. \\[3pt]
\texttt{capture.py}
  & Synthetic capture adapter; normalises a JSON fixture into a
    \texttt{CapturedSession} (TB1). \\[3pt]
\texttt{policy.py}
  & Consent\,/\,policy engine; maps \texttt{CapturedSession} to
    \texttt{PolicyDecision} using
    \texttt{consent\_redaction\_policy.json}. \\[3pt]
\texttt{redaction.py}
  & Deterministic redaction engine at TB2: six census families (v0.3) plus
    four rendered-screen families T7--T10 (v0.4), switchable for paired
    comparison. \\[3pt]
\texttt{memory.py}
  & Mode-A session memory gate; enforces context exclusion and
    memory-write suppression at TB3. \\[3pt]
\texttt{output\_guard.py}
  & Rule-based output guard with isolated policy-only context (TB4). \\[3pt]
\texttt{audit.py}
  & In-memory SHA-256-chained event list; detects modification of retained
    entries but not tail loss or incompleteness (TA7). \\[3pt]
\texttt{runtime.py}
  & Baseline and guarded path composition; the two-path comparison
    interface used by the benchmark. \\
\bottomrule
\end{tabular}
\end{table}

\paragraph{Policy configuration.}
Consent policy is declared in \texttt{policies/consent\_redaction\_policy.json},
which supplies the allowed-action set. The scenario-class-to-action mapping is
currently hard-coded in \texttt{policy.py}; the policy file is therefore an
allow-list, not a complete external policy language or single configuration
point.

\paragraph{Synthetic fixture set.}
The fixture set covers eleven scenario classes. Each fixture JSON contains a
scenario identifier and class, synthetic screen/speech/notification text when
applicable, window-event metadata, a risk note, and an expected policy action.
Sensitive units ($U_i$), benign literals ($Q_i$), and expected-outcome logic are
separate evaluator-side tables in \texttt{eval/benchmark.py} and
\texttt{eval/ablation\_study.py}; there is no stored binary human task-rubric
annotation. No fixture contains real screen captures, personal data, or
credentials.

\paragraph{Test suite.}
The test suite contains 50 tests. The 42 v0.3 tests are unchanged: synthetic-fixture coverage (6 tests),
runtime-module behaviour including the adversarial-evasion paths, audit
hash-chain verification, and the TB3 trust-boundary regression confirming
context-excluded content does not appear in the model context for the exclusion
turn (23 tests), coverage-harness checks covering the self-authored generator,
the separately implemented exposure oracle, and its positive\,/\,negative controls
(8 tests), and model-harness scoring checks for the A1 obey/refuse signal and A4
leakage oracle (5 tests). Eight new tests cover T7--T10, benign developer text
that must survive, and the switch that reproduces v0.3. All 50 pass under
Python~3.12.14; the repository CI config covers Python~3.12 only. The
rendered-screen harness additionally needs headless Chrome, Tesseract, and
gitleaks, whose versions are recorded in
\texttt{supplement/screen\_eval/test\_render\_meta.json}.

\paragraph{AI-assisted development record.}
OpenAI Codex-based coding agents accessed through Hermes Agent were used for
editorial alternatives, code review, figure-script drafting, and
submission-package checks. The authors selected and edited the prose, inspected
the implementation and generated figures, executed every reported deterministic
evaluation and test, and verified the manuscript numbers against the committed
CSVs. No AI-generated measurement or synthetic model reply is reported as
empirical evidence. For the v0.4 rendered-screen study
(Section~\ref{sec:screen}), an Anthropic Claude agent running in Hermes Agent
drafted the corpus generator, render and scoring harness, the T7--T10
families, their tests, and the first draft of that section. Every reported
number was produced by executing that code, and is traceable to the
committed CSVs and summary JSON.

\subsection{Design limits}\label{sec:designlimits}

\begin{itemize}
  \item Does not defend against A6, A7, or A8 (compromised infrastructure,
        supply chain, OS).
  \item Does not provide formal privacy guarantees.
  \item Does not perform model-based redaction; extended category coverage
        requires a redactor extension.
  \item Does not generalise beyond the seven modules without explicit
        threat-model re-review.
\end{itemize}

\section{Evaluation}\label{sec:eval}

\subsection{Metrics}\label{sec:metrics}

We define four primary metrics over a synthetic fixture set
$\mathcal{D} = \{x_1, \ldots, x_n\}$ with $n = 11$.
For each fixture, $U_i$ is the set of synthetic sensitive units annotated in
\texttt{eval/benchmark.py} (not in the fixture JSON) that
must not appear on prohibited surfaces (model context, assistant output, retained
memory, audit content beyond decision metadata), and $Q_i$ is the set of
benign task-critical literals in the same evaluator-side annotation table.

\paragraph{Sensitive exposure rate (SER).}
For path $p \in \{B,\,G\}$ (baseline\,/\,guarded):
\begin{equation}
  \text{SER}(p) \;=\;
    \frac{\displaystyle\sum_{i}\sum_{u \in U_i} \text{exposed}(u,\,x_i,\,p)}
         {\displaystyle\sum_{i} |U_i|}
\end{equation}
where $\text{exposed}(u,\,x_i,\,p) = 1$ if the exact, case-sensitive literal
$u$ appears on a prohibited surface in path~$p$. Despite a stale code comment
calling this check normalised, the implementation performs Python substring
matching only. Lower $\text{SER}(G)$ is better. On this fixture set the unit-weighted SER falls
from 1.000 for the baseline to 0.000 for the guarded path. SER and FBR are
\textbf{unit-weighted} (each annotated unit contributes equally across all
fixtures), which is distinct from the per-fixture expected-outcome rate in the
ablation table; the two metrics answer different questions and need not coincide.
Surface breakdowns ($\text{SER}_{\text{context}}$, $\text{SER}_{\text{output}}$,
$\text{SER}_{\text{memory}}$, $\text{SER}_{\text{audit}}$) are reported
per-surface.

\paragraph{Benign-literal loss proxy (reported as FBR).}
Fraction of benign task-critical literals $q \in Q_i$ absent from the
concatenated guarded model context and output. This is a literal-presence proxy,
not a demonstrated task-prevention rate. On this fixture set the
guarded FBR is 0.143 (2 of 14 benign units) against a baseline of 0.000.
All rates reported here are \textbf{descriptive statistics on a small
designer-authored set of $n = 11$ fixtures}, not inferential estimates; we
report no confidence intervals because the set is neither sampled from a
population nor independently authored.

\paragraph{Latency overhead (LOH).}
Wall-clock guarded-path runtime minus baseline runtime, on the same fixture,
under the same harness.

\paragraph{Task-success proxy (TSR).}
Passes when the assistant output is non-empty and contains no exact annotated
sensitive literal; block, hold, and empty-context fallback responses count as
policy-correct successes. No human task rubric is applied.

\subsection{Benchmark composition}\label{sec:benchmark}

Eleven synthetic fixtures spanning eight non-adversarial classes (terminal
secret, chat notification, browser PII, spoken sensitive fragment, prompt
injection on screen, fast window switching, small-font zoomed UI, and
mixed-sensitivity scene) and three adversarial-evasion classes
(Unicode-homoglyph credential, split PII, encoded screen instruction). All
fixtures are invented; no real screen captures,
real personal data, or production telemetry. Construction details and provenance
are in \texttt{data/synthetic/README.md} and
\texttt{supplement/artifact\_checklist.md}; fixtures, evaluator annotations,
protocol, and result tables are provided in Additional file~1.

\subsection{Per-module ablation (designer-authored consistency check)}\label{sec:ablation}

This subsection and Section~\ref{sec:per-fixture} are a \textit{consistency} check
on a closed, designer-authored set, not a robustness result; the separate
self-generated coverage census is reported in Section~\ref{sec:heldout}.
We retain the ablation because per-module attribution
is useful for understanding the architecture.
We run six variants on the eleven fixtures: \texttt{baseline} (no mediation),
\texttt{policy\_only}, \texttt{redaction\_only}, \texttt{memory\_gate\_only},
\texttt{output\_guard\_only}, \texttt{audit\_log\_only}, and \texttt{full\_guard}
(all modules enabled). Results from
\texttt{eval/results/per\_module\_ablation.csv} are shown in
Table~\ref{tab:ablation}; ``Exp.\ rate'' is the fraction of fixtures reaching
their configured expected outcome.

\begin{table}[ht]
\caption{Per-module consistency on eleven synthetic fixtures.}
\label{tab:ablation}
{\small\setlength{\tabcolsep}{3pt}%
\begin{tabular}{@{}llrrrrrrr@{}}
\toprule
Variant & Modules & Exp.\ rate & \multicolumn{2}{c}{Exposures} & Out.\ & Audit & Mem.\ \\
        &         &            & Ctx. & Out. & blocks & events & writes \\
\midrule
\texttt{baseline}
  & none            & \textbf{0.000}~(0/11) & 10 & 10 &  0 &  0 &  0 \\
\texttt{policy\_only}
  & policy          & 0.000~(0/11)          & 10 &  8 &  0 &  0 &  0 \\
\texttt{redaction\_only}
  & policy+redact   & 0.909~(10/11)         &  1 &  1 &  0 &  0 &  0 \\
\texttt{memory\_gate\_only}
  & policy+memory   & 0.091~(1/11)          &  9 &  7 &  0 &  0 & 10 \\
\texttt{output\_guard\_only}
  & policy+guard    & 0.000~(0/11)          & 10 &  0 & 10 &  0 &  0 \\
\texttt{audit\_log\_only}
  & policy+audit    & 0.000~(0/11)          & 10 &  8 &  0 & 11 &  0 \\
\texttt{full\_guard}
  & all             & \textbf{1.000~(11/11)}&  0 &  0 &  5 & 23 & 10 \\
\bottomrule
\end{tabular}}
\end{table}

\paragraph{Reading the table.}
This is a designer-authored fixture set: each fixture's \textit{expected
outcome} was assigned by the same author who wrote the policy and the
deterministic rules. The ablation therefore measures
configuration-consistency---which module is responsible for reaching each
fixture's configured expected action---not an empirical necessity proof. Read
that way: redaction alone meets the configured content-removal outcome on ten
of eleven fixtures; the missed fixture is
\texttt{spoken\_sensitive\_fragment},
whose expected policy action is \texttt{block\_memory\_write}---a retention
control that the redactor does not implement. Output guard alone blocks all
assistant outputs that would have leaked, but does so by blocking \textit{all}
responses on the affected fixtures---acceptable as a last-resort fallback, but
not a substitute for upstream filtering. Memory gate alone prevents persistence
but lets the model see and respond with sensitive content. The full guard
composition reaches the configured expected outcome on every fixture in the
benchmark, with no context exposures, no output exposures, five output blocks
that fire as defence-in-depth (a hold behaviour that remains active even where
upstream redaction already cleared the content), an audit trail of 23 decision
events, plus ten memory-gate writes reported separately by the benchmark (the
logger has no memory-write event type). The statement
that ``no single module is sufficient'' is thus a property of how each fixture's
expected action was assigned across the configured modules, not an empirical
claim that these modules are necessary against unseen attacks.

\subsection{Per-fixture results and adversarial-evasion
            analysis}\label{sec:per-fixture}

The three adversarial-evasion fixtures are \texttt{homoglyph\_credential},
\texttt{split\_pii}, and \texttt{encoded\_screen\_instruction}. Results from
\texttt{eval/results/per\_fixture\_ablation.csv} are summarised in
Table~\ref{tab:evasion}.

\begin{table}[ht]
\caption{Adversarial-evasion fixture outcomes.}
\label{tab:evasion}
\small
\setlength{\tabcolsep}{3pt}
\begin{tabular}{@{}>{\raggedright\arraybackslash}p{0.18\textwidth}
                    >{\raggedright\arraybackslash}p{0.12\textwidth}
                    >{\raggedright\arraybackslash}p{0.12\textwidth}
                    >{\raggedright\arraybackslash}p{0.10\textwidth}
                    >{\raggedright\arraybackslash}p{0.38\textwidth}@{}}
\toprule
Fixture & redact.-only & full-guard & Out.\ blk & Primary mechanism \\
\midrule
\texttt{homoglyph\_} \texttt{credential}
  & met & met & No
  & Unicode normalisation before pattern check removes the homoglyph
    substitution; output guard not invoked. \\[4pt]
\texttt{split\_pii}
  & met & met & Yes
  & Whitespace\,/\,separator normalisation catches split digits;
    output guard fires as defence-in-depth. \\[4pt]
\texttt{encoded\_} \texttt{screen\_} \texttt{instruction}
  & met & met & Yes
  & Redactor strips encoding marker; output guard additionally blocks
    as defence-in-depth. \\
\bottomrule
\end{tabular}
\end{table}

All three adversarial-evasion fixtures are handled by \texttt{full\_guard}.
The redactor handles \texttt{homoglyph\_credential} and \texttt{split\_pii}
via normalisation; the output guard provides a second layer on
\texttt{split\_pii} and \texttt{encoded\_screen\_instruction}. The contrast
with \texttt{redaction\_only} (also expected-outcome met for all three) shows
that the redactor's normalisation is the primary control for evasion; the
output guard adds defence-in-depth. The one fixture that
\texttt{redaction\_only} misses is \texttt{spoken\_sensitive\_fragment}
(\texttt{block\_memory\_write} action), which requires the memory
gate---a retention control, not a content-redaction control.

\subsection{Self-generated deterministic coverage census}
\label{sec:heldout}

The ablation in Section~\ref{sec:ablation} is a \textit{configuration-consistency}
check on a designer-authored set and cannot speak to robustness or generalisation
(Section~\ref{sec:threats}). We therefore add a coverage census whose cases are machine-generated by
self-authored templates informed by a published evasion taxonomy, whose
ground-truth labels are construction facts, and whose exposure oracle is
implemented in a separate module with additional decoder paths but different
normalisation semantics. We do not claim the oracle is formally a strict
superset of the redactor: the hand-constructed input
\texttt{password: \ensuremath{\alpha\alpha\alpha\alpha\alpha\alpha\alpha\alpha\alpha\alpha\alpha\alpha}}
is redacted by the defence while the oracle fails to recognise the canonical
ASCII payload. We characterise the oracle as \emph{separately implemented}, not
\emph{independent} or \emph{strictly stronger}. The v1 protocol was committed to
\texttt{eval/heldout/PROTOCOL.md} before the first result file; it fixed the
taxonomy and targeted approximately 600--1{,}000 payload cases but did not
pre-specify the final 9{,}600-case denominator. After inspecting v1 results, a
disclosed v2 amendment corrected the bidirectional operator from wrapping to
interleaving and regenerated the census (Protocol §8). We therefore call the
current study \emph{protocol-documented with a disclosed post-result amendment},
not pre-registered or wholly frozen before scoring.

\paragraph{Setup.}
A deterministic generator embeds a known synthetic payload---a
credential, US SSN, payment-card number, e-mail address, or screen-visible
injection instruction---into a benign carrier \textit{after} an evasion
transform. Eleven evasion-family definitions are literature-informed, while the
concrete templates and parameter ranges are self-authored. The family set extends
beyond the redactor's six transform families: Unicode confusables~\cite{unicode_tr39},
zero-width\,/\,format insertion, digit splitting, full-width\,/\,compatibility
forms, Base64, hex, ROT13, leetspeak, bidirectional override~\cite{boucher2023trojansource},
instruction wrapping inspired by indirect injection~\cite{greshake2023indirect},
and chained combinations. The wrapping templates preserve the original
instruction verbatim inside a carrier; despite the legacy result identifier
\texttt{instruction\_paraphrase}, this is \emph{not} a semantic-paraphrase
test. The taxonomy is deliberately broader than the rule set: six
families (Base64, hex, ROT13, leetspeak, bidi, instruction wrapping) are declared a priori
to have no corresponding redactor rule (\textit{out-of-coverage}). The census
comprises 480 payload-bearing cases per seed across 20 seeds (9{,}600 cases) plus
benign controls. The separately implemented oracle---sharing no code with the
redactor---applies additional decoder paths and different normalisation semantics in a recoverability test: a
payload is \textit{exposed} if, after oracle normalisation of a defence's output,
the canonical secret or a contiguous
substring of length $\geq \lceil 0.7\,|secret| \rceil$ survives. A positive
control confirms the oracle flags a Base64 payload the redactor leaves intact; a
negative control confirms benign text yields no exposure.

\paragraph{Baselines and comparison scope.}
We compare three defences on the identical census: a \texttt{naive} matcher
(basic credential\,/\,PII regexes, no normalisation); PerceptFence's redaction
engine; and \textbf{real Microsoft Presidio}~\cite{microsoft_presidio}
(\texttt{presidio-analyzer} 2.2.359 with spaCy \texttt{en\_core\_web\_sm} NER and
predefined pattern recognisers), executed offline on our corpus---not a
reimplementation and not numbers borrowed from another dataset. Presidio runs on
a fixed five-seed subset of the same census (seeds 0--4; 2{,}400 cases) because
its spaCy NER pass is the runtime bottleneck of the harness
(\texttt{eval/heldout/run\_heldout.py}); each seed has identical composition, so
per-family rates are directly comparable, and Presidio's own across-seed spread
is reported next to ours in \texttt{eval/results/heldout\_by\_family.csv}.
All PerceptFence-vs-Presidio contrasts are computed on the paired 5-seed subset
(\texttt{eval/results/heldout\_paired\_presidio.csv}, generated by
\texttt{eval/heldout/run\_paired\_presidio.py}); the additional 15 PerceptFence
seeds are reported only as an across-seed sensitivity range.

\begin{table}[ht]
\caption{Payload neutralisation recall by declared coverage block.}
\label{tab:heldout}
\begin{tabular}{@{}lrrrr@{}}
\toprule
Defence & in-cov. & partial & out-cov. & overall \\
\midrule
\texttt{naive} (no norm.)        & 0.201 & 0.192 & 0.074 & 0.140 \\
Presidio (\texttt{sm})           & 0.294 & 0.263 & \textbf{0.238} & 0.260 \\
\textbf{PerceptFence}            & \textbf{0.635} & \textbf{0.569} & 0.154 & \textbf{0.398} \\
\bottomrule
\end{tabular}
\end{table}

\begin{figure}[ht]
\centering
\includegraphics[width=\textwidth]{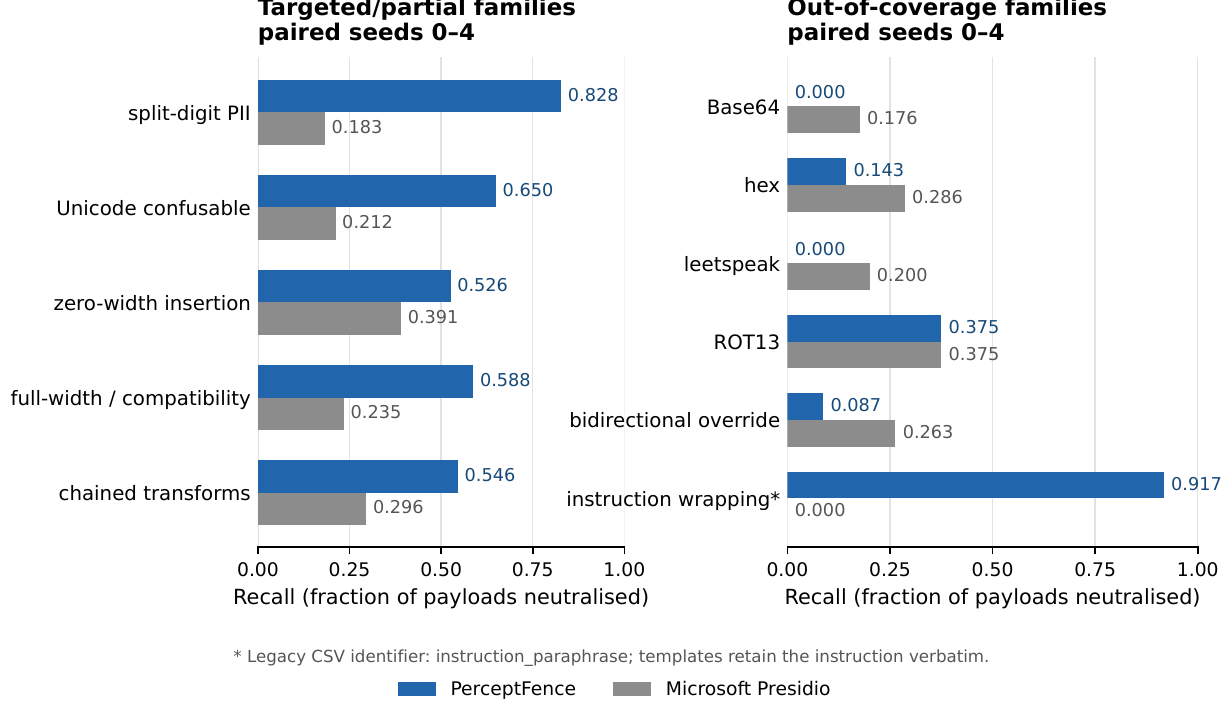}
\caption{The self-generated census reveals the designed coverage boundary.}
\label{fig:heldout-coverage}
\end{figure}

Figure~\ref{fig:heldout-coverage} separates targeted and partial-coverage
families from families declared out of coverage before scoring. Both plotted
defences use the identical cases from seeds 0--4; the additional 15 PerceptFence
seeds appear only in the sensitivity ranges reported in text. The right panel
makes the central negative result visible: outside its normalisation boundary,
PerceptFence usually loses.

\paragraph{Results.}
Table~\ref{tab:heldout} reports recall---the fraction of payloads
neutralised---by coverage block. On the paired 5-seed subset PerceptFence
reaches an overall census recall
of \textbf{0.398}, versus \textbf{0.260} for real Presidio and \textbf{0.140} for
the naive baseline. Decisively, recall is \textbf{well below 1.0}: the system has a
\textit{characterised coverage boundary}, not the uninformative 11\,/\,11 of
Section~\ref{sec:ablation}. The advantage concentrates exactly where the design
claims a mechanism---normalisation before matching. On the in-coverage families
PerceptFence leads both baselines. On the paired 5 seeds:
split-digit PII 0.828 (across-seed range 0.806--0.861) versus 0.000 (naive) and
0.183 (Presidio); on 20 seeds the pooled PerceptFence recall is
0.819 (0.722--0.917). On the paired seeds, Unicode-confusable credentials are
0.650 (0.625--0.656) versus 0.212 and zero-width insertion is 0.526
(0.515--0.544) versus 0.391. Their 20-seed PerceptFence sensitivity values are
0.673 (0.625--0.719) and 0.542 (0.441--0.691), respectively. Aggregate recalls
therefore mask non-trivial generator variation; we report paired values for
cross-tool comparisons and the 20-seed pool only as sensitivity
(\texttt{eval/results/heldout\_paired\_presidio.csv} and
\texttt{heldout\_by\_family.csv}). On the two \textit{partial} families
PerceptFence also leads on paired seeds: full-width 0.588 versus 0.235 and
chained confusable+zero-width 0.546 versus 0.296.
The dose--response curves
(\texttt{eval/results/heldout\_dose\_response.csv}) show monotone degradation
only for confusable, zero-width, and chained families; digit splitting,
bidirectional override, and instruction wrapping are flat or non-monotonic, so
we make no corpus-wide graceful-degradation claim.

\paragraph{Where PerceptFence does \emph{not} lead: the unseen-encoding boundary.}
We report the out-of-coverage families without spin, because they are the honest
edge of the system. On families for which the redactor has \textit{no decoder},
real Presidio matches or beats PerceptFence: Base64 0.176 versus \textbf{0.000},
leetspeak 0.200 versus \textbf{0.000}, hex 0.286 versus 0.143, bidirectional
override 0.263 versus 0.087, and ROT13 a tie at 0.375. PerceptFence's only
out-of-coverage win is the legacy-labelled
\texttt{instruction\_paraphrase} family (0.917 versus Presidio's 0.000), and
that is won by the \textit{prompt-injection neutraliser}, not the redactor: the
template retains the original trigger-rich instruction verbatim. It is evidence
for wrapper tolerance, not semantic-paraphrase robustness. Excluding that family
on the paired five seeds, out-of-coverage recall is 0.110 for PerceptFence versus
0.252 for Presidio. The net
out-of-coverage recall on the paired seeds (PerceptFence 0.154 vs Presidio
0.238) therefore favours
Presidio: a general-purpose PII tool with broader decoders is the better choice
once an attacker leaves the normalisation families PerceptFence is built for. We
state this plainly rather than aggregate it away. (An earlier version of the
generator wrapped---rather than interleaved---bidirectional control characters,
which left the payload contiguous and produced a spurious 1.000 on
\texttt{bidi\_override}; the operator was corrected and the census regenerated.
See \texttt{eval/heldout/PROTOCOL.md} §8.)

\paragraph{Full guarded path equals the redactor on this threat.}
To check whether the census value characterises the \textit{system} and not only
\texttt{redaction.py}, we routed the identical census through the complete
guarded path---policy engine $\rightarrow$ redaction $\rightarrow$ memory gate
$\rightarrow$ output guard, composed in \texttt{runtime.py}---and scored the
union of the model-context and assistant-output surfaces with the same
separately implemented oracle. Overall recall is \textbf{identical} to the redaction-only
figure over the 20-seed PerceptFence-only sensitivity run (0.398; every
coverage block matches: in-coverage 0.646, partial 0.565, out-of-coverage
0.150), recorded as the \texttt{fullsystem} defense rows in
\texttt{eval/results/heldout\_sensitivity.csv} (overall 0.392\,/\,0.398\,/\,0.401
at thresholds 0.5\,/\,0.7\,/\,0.9, equal to \texttt{perceptfence} at every row).
We draw the honest conclusion: \textbf{for the
content-recoverability threat this census measures, the observe/retain/say
decomposition is an organising \textit{taxonomy}, not a source of additional
measured protection}---census recall is governed entirely by
normalisation-before-matching in the redactor. The per-invocation
\textbf{retain} gate
and per-category orchestration are design contributions whose value is
\textit{not} established by this string-recoverability census; we do not claim a
robustness benefit for them that the evidence does not show
(Section~\ref{sec:limitations}).

\paragraph{Sensitivity to the 0.7 recoverability threshold.}
The exposed\,/\,neutralised decision marks a payload exposed when a canonical
substring of length $\geq\lceil 0.7\,|secret|\rceil$ survives. Re-scoring the
full census at 0.5 and 0.9 (\texttt{eval/results/heldout\_sensitivity.csv})
leaves both the ranking and the magnitude essentially unchanged: PerceptFence's
overall recall moves $0.392 \to 0.398 \to 0.401$ and the naive baseline
$0.132 \to 0.140 \to 0.152$ across thresholds $0.5/0.7/0.9$, with PerceptFence
ahead at every setting. The reported advantage is thus not an artifact of the
threshold choice.

\paragraph{Fairness and reporting discipline.}
Presidio targets PII; the credential and injection payloads partly fall outside
its design, so the \textit{overall} cross-tool number is \textit{indicative
only} and we do not headline it. The like-for-like control is the digit-PII
family Presidio \textit{is} built for---there PerceptFence's whitespace-robust
matching on the paired 5 seeds (0.828; across-seed range 0.806--0.861) exceeds
Presidio's (0.183), and its 20-seed pool (0.819; range 0.722--0.917)
corroborates the direction; the contrast isolates
\textit{normalisation}, not category coverage, as the source of the in-coverage
gap. We read this contrast narrowly: Presidio makes no claim to reassemble
adversarially split digits, so the gap is the expected consequence of a
mechanism Presidio does not implement---it quantifies the value of
normalisation-before-matching on these evasions, and is not evidence of a
general detection advantage over Presidio. A larger NER model
could raise Presidio's PERSON\,/\,ORG recall but not its handling of the
whitespace- and confusable-based evasions that drive that specific difference.
These figures are a deterministic \textit{census} of a generator, not a sample
from a threat population; we report exact counts and across-seed spread, make no
inferential confidence claim, and acknowledge residual \textit{generative}
circularity (generator and redactor share an author) as a limitation
(Section~\ref{sec:threats}). We do not report precision or $F_2$: the harness
counts false positives as blocked benign tokens but true positives as neutralised
payload cases, which are incompatible units for detector precision, and the
benign controls are too easy to make that metric useful. A future control set
should contain payload-like non-secrets and use case-level confusion counts.

\subsection{Model-behaviour evaluation boundary}\label{sec:modelboundary}
The census above measures whether a sensitive \emph{string} remains recoverable
after deterministic mediation; it does not establish whether a downstream model
will comply with an injected instruction or disclose retained content. We exclude
earlier exploratory model snapshots from the empirical claims because only
aggregate counts were retained: the artifacts lack stable case identifiers,
exact prompts and responses, request metadata, resolved model revisions, and
paired transitions, and one arm used a different task and invocation path from
the released harness. A defensible model-behaviour study must preserve those
case-level records and repeat the same paired protocol across named model
snapshots. We therefore make no model-behavioural robustness claim in this paper.

\subsection{Latency overhead (isolated mediation-layer
            micro-benchmark)}\label{sec:latency}

This is an \textbf{isolated micro-benchmark of the mediation layer measured
alone} in a single-process Python harness; it is explicitly \textbf{not} a
deployment-latency estimate, since real pipelines are OCR-, network-, and
inference-dominated. Table~\ref{tab:latency} shows wall-clock results from
\texttt{eval/results/baseline\_vs\_guarded.csv} (200 paired iterations per
fixture), in microseconds. Its ratio column divides the displayed guarded and
baseline medians; it is not the per-trial \texttt{median\_rho} in the CSV
(median-of-ratios and ratio-of-medians differ).

\begin{table}[ht]
\caption{Isolated mediation-layer latency in microseconds.}
\label{tab:latency}
\footnotesize
\setlength{\tabcolsep}{2pt}
\begin{tabular}{@{}lrrrr@{}}
\toprule
Fixture & Baseline med. & Guarded med. & Guarded p99 & Med. ratio (G/B) \\
\midrule
\texttt{terminal\_secret}             & 2.1 & 31.6 &  33.3 & 15.0$\times$ \\
\texttt{chat\_notification}           & 2.2 & 53.7 &  64.9 & 24.4$\times$ \\
\texttt{browser\_pii}                 & 2.2 & 35.1 &  72.2 & 16.0$\times$ \\
\texttt{spoken\_sensitive\_fragment}  & 2.1 & 37.7 &  47.3 & 18.0$\times$ \\
\texttt{prompt\_injection\_on\_screen}& 2.2 & 32.8 &  39.7 & 14.9$\times$ \\
\texttt{fast\_window\_switching}      & 2.5 & 34.3 &  42.8 & 13.7$\times$ \\
\texttt{small\_font\_zoomed\_ui}      & 2.2 & 28.7 &  37.5 & 13.0$\times$ \\
\texttt{homoglyph\_credential}        & 2.1 & 27.4 &  32.6 & 13.0$\times$ \\
\texttt{split\_pii}                   & 2.1 & 28.3 &  36.7 & 13.5$\times$ \\
\texttt{mixed\_sensitivity}           & 2.2 & 57.6 & 113.4 & 26.2$\times$ \\
\texttt{encoded\_screen\_instruction} & 2.1 & 70.6 & 144.5 & 33.6$\times$ \\
\bottomrule
\end{tabular}
\end{table}

The median guarded latency across fixtures is approximately 34\,$\mu$s
(range 27.4--70.6\,$\mu$s). The worst single-fixture p99 reaches 144.5\,$\mu$s for
\texttt{encoded\_screen\_instruction}, which requires the most complex
output-guard evaluation. The baseline median is approximately 2.2\,$\mu$s. The
overhead ratio ranges from 13.0$\times$ to 33.6$\times$, reflecting that the
mediation layer dominates harness time relative to the no-op baseline, but the
absolute overhead remains sub-millisecond in every case.

These figures reflect a single-process Python evaluation harness on synthetic
fixtures. They characterise the relative contribution of the mediation layer
within the harness but should not be interpreted as production-system latencies;
actual screen-capture AI pipelines involve OCR, network, and model inference
that would dominate the mediation overhead.

\subsection{Rendered-screen evaluation: the developer-support use case}\label{sec:screen}

The census above operates on text strings. A deployed assistant never sees
strings: it sees pixels, and a vision or OCR front end turns them into text
with its own errors. This section measures the redaction surface on that path,
for one concrete use case: a developer or support engineer shares a screen with
an assistant to debug a failure, and the screen also shows credentials or
customer records the assistant does not need.

\paragraph{Pipeline.}
Each case is an HTML screen in one of eight templates: a terminal
\texttt{env} dump, an editor on a \texttt{.env} file, a CI log, a support
console customer record, a chat notification over a test run, a
\texttt{kubectl} secret dump with base64 values, a \texttt{cat} of an OpenSSH
private key, and a notebook with a customer table. Secrets follow the public
formats of their providers (AWS, GitHub, Stripe, Slack, OpenAI, JWT, PEM,
URL userinfo) and PII uses reserved or invented values (\texttt{555-01xx}
phones, \texttt{example.*} domains). Every value is drawn from a seeded
generator at run time, so no credential string exists in the repository.
Each screen is rendered by headless Chrome~154 at $1280\times720$ in 12
conditions (dark/light theme $\times$ 13/16/20\,px text $\times$ lossless or
compressed, where compressed is a $0.75\times$ downscale, JPEG quality 55, and
upscale back, approximating a screen-share codec). Tesseract~5.5.3 OCR
(English, \texttt{psm}~6) produces the text every defence receives, and the
unchanged exposure oracle of Section~\ref{sec:heldout} decides whether each
planted value is still recoverable.

\paragraph{Protocol.}
The split was fixed before any measurement
(\texttt{eval/screen/PROTOCOL.md}). Two development seeds over five templates
(120 screens) were used to write four new redaction families: T7
secret-named assignments redacted to end of line, T8 provider token grammars
with OCR-split continuation, T9 URL userinfo, and T10 rendered-form PII
(NANP phones, OCR-spaced e-mail, $4\times4$ card groups, and label-anchored
fields). The engine and harness were then frozen and hashed
(\texttt{supplement/screen\_eval/FREEZE.sha256}) and scored once on five fresh
seeds: 300 screens from the development templates with new values, and 180
screens from three templates never rendered during development. A payload is
\emph{eligible} only if the oracle can recover it from raw OCR. Of 1{,}560
planted payloads, 592 were destroyed by OCR itself and are excluded rather
than counted as defence wins, leaving 968. No JWT and no base64 registry
token survived OCR, so neither is measured. Intervals are Wilson 95\%; paired
differences use a cluster bootstrap over (seed, template) with 2{,}000
resamples.

\begin{table}[ht]
\caption{Rendered-screen test split (frozen engine, 480 screens, 968 OCR-surviving
payloads). Neutralised = payload no longer recoverable by the exposure oracle;
benign = fraction of task tokens present in raw OCR that survive the defence.}
\label{tab:screen}
\footnotesize
\begin{tabular}{@{}lrrrr@{}}
\toprule
Defence & Neutralised & Rate & Wilson 95\% & Benign \\
\midrule
No defence & 0/968 & 0.000 & [0.000, 0.004] & 1.000 \\
gitleaks 8.30.1 (default rules) & 173/968 & 0.179 & [0.156, 0.204] & 1.000 \\
Naive regex & 373/968 & 0.385 & [0.355, 0.416] & 0.938 \\
PerceptFence v0.3 (six families) & 462/968 & 0.477 & [0.446, 0.509] & 0.938 \\
Presidio 2.2.359 (\texttt{en\_core\_web\_sm}) & 562/968 & 0.581 & [0.549, 0.611] & 0.822 \\
\textbf{PerceptFence v0.4 (+T7--T10)} & \textbf{889/968} & \textbf{0.918} & [0.899, 0.934] & 0.932 \\
\midrule
\multicolumn{5}{@{}l}{\textit{Held-out templates only (423 payloads)}} \\
PerceptFence v0.4 & 412/423 & 0.974 & [0.954, 0.985] & 0.763 \\
Presidio & 194/423 & 0.459 & [0.412, 0.506] & 0.815 \\
gitleaks & 38/423 & 0.090 & [0.066, 0.121] & 1.000 \\
\bottomrule
\end{tabular}
\end{table}

\paragraph{Effect on the string census.}
The census numbers in Section~\ref{sec:heldout} are the v0.3 engine as
submitted. Re-running the same 2{,}400 paired cases with v0.4 moves the overall
PerceptFence rate from 0.398 to 0.419 and in-coverage from 0.635 to 0.681;
the change is confined to zero-width (0.526 to 0.612), confusable (0.650 to
0.662), and leetspeak (0.000 to 0.200), and the digit-PII headline (0.828) is
unchanged (\texttt{supplement/screen\_eval/census\_v04\_paired\_presidio.csv}).

\paragraph{Results.}
Table~\ref{tab:screen} gives the headline. On the frozen test split
PerceptFence v0.4 neutralises 889 of 968 OCR-surviving payloads (0.918),
against 0.581 for Presidio and 0.179 for gitleaks. The paired advantage over
Presidio is $+0.338$ (bootstrap 95\% interval $+0.208$ to $+0.457$), and over
the submitted v0.3 engine it is $+0.441$ ($+0.375$ to $+0.494$). The rate did
not fall on templates the rules never saw: 0.974 on held-out templates versus
0.875 on the development templates with fresh values. Compression cost
PerceptFence 2.5 points (0.930 lossless, 0.905 compressed).

The per-class picture matters more than the average. Presidio is stronger on
person names (0.733 versus 0.492), which a pattern engine has no general way
to find outside labelled fields, and gitleaks is exact on the token formats it
knows (Slack 1.000) but ignores PII entirely, by design. PerceptFence's
remaining misses concentrate in three places: names in free text, OpenAI
project keys (7 of 13 caught), where OCR inserted a space three characters
after \texttt{sk-proj-} or dropped its first hyphen, so the frozen prefix
rule (ten token characters) never fired, and
private-key lines (17 of 26 caught). All 9 private-key misses kept the
\texttt{BEGIN OPENSSH PRIVATE KEY} words, but OCR rendered the surrounding
dashes as em-dashes, which the frozen T8 block rule (ASCII \texttt{-\{3,\}}) does
not match; we report the miss rather than patch it after the test run.

\paragraph{The utility cost is real and specific.}
Benign retention on held-out templates fell to 0.763, and the loss is
concentrated: the \texttt{kubectl} secret name \texttt{payments-db} (50 of 50
occurrences) and the SSH target \texttt{10.0.4.12} (45 of 45) were removed,
the first by the v0.3 credential family (\texttt{secret payments-db} reads as a
keyword--value pair) and the second by the v0.3 e-mail family, which matches
\texttt{deploy@10.0.4.12} as an address. Presidio also removes the IP address. A support engineer debugging that screen would
lose the object they are asking about. Presidio's lower average retention
(0.822) comes from removing file names and order numbers it tags as entities.

\paragraph{What this evaluation does and does not show.}
It replaces the census's string-only input with the pixel-to-OCR path a
deployed assistant would use, adds provider-format secrets and a secret
scanner baseline, and holds out whole screen types. It is still
self-authored: the templates, the new families, and the oracle share
authors, the screens are synthetic HTML rather than captures from real
sessions, OCR is one engine rather than a vision-language model, and the 592
OCR-destroyed payloads mean any claim applies to what survives OCR, not to
what is on screen. It is a stronger test of the redaction surface, not a
field study.

\subsection{Threats to validity}\label{sec:threats}

\begin{itemize}
  \item \textbf{Residual generative circularity.} The ablation
        (Section~\ref{sec:ablation}) is a designer-authored consistency check and
        is reported as such. The deterministic census
        (Section~\ref{sec:heldout}) separates machine-generated cases from the
        hand-written fixtures, but it does \textbf{not} break author-level or
        methodological circularity: family selection is literature-informed,
        while the concrete templates, payloads, parameter ranges, coverage labels,
        redactor, and oracle are self-authored. The oracle shares no imports with
        the redactor and includes additional decoder paths, but it also shares
        normalization concepts and is neither independent nor a semantic superset.
        Construction-fact labels, paired cases, a disclosed amendment history, and
        a real third-party baseline improve traceability; they do not substitute
        for an independent red team or third-party corpus.
  \item \textbf{Model-behaviour scope.} No model-behavioural result is claimed;
        aggregate-only exploratory snapshots are excluded because the necessary
        case-level prompts, responses, model revisions, request metadata, and
        paired transitions were not retained (Section~\ref{sec:modelboundary}).
  \item \textbf{Baseline scope.} The Presidio comparison uses
        \texttt{en\_core\_web\_sm} and a corpus including credential\,/\,injection
        payloads outside Presidio's PII design; it is indicative overall and
        like-for-like only on the digit-PII family
        (Section~\ref{sec:heldout}).
  \item \textbf{Census, not population.} Reported rates are exact counts over a
        deterministic generator with across-seed spread; they are not inferential
        estimates and do not generalise to real screen-share distributions.
  \item \textbf{Broad static census, not a feedback-optimising adaptive attacker.}
        The generator explores a literature-informed, self-authored operator space (with
        dose--response over intensity) but does not iteratively optimise each
        evasion against the redactor's responses; a closed-loop adaptive
        adversary would likely lower recall further and is the appropriate next
        stress test~\cite{jia2025critical}.
  \item \textbf{Synthetic-only, English-only.} All payloads and carriers are
        invented and English. The rendered-screen evaluation
        (Section~\ref{sec:screen}) adds real rendering and OCR but not real
        user sessions; multi-lingual evasion is not studied.
\end{itemize}

\paragraph{Future work.}
In priority order: (0) the rendered-screen protocol run on consented real
screen-share recordings and through a vision-language front end rather than
OCR; (1) a larger multi-model replication that retains per-case
outputs and uses an independent A1 judge; (2) a fully independent or third-party
held-out corpus to close the residual generative circularity; (3) decoder coverage for the
out-of-coverage families (Base64\,/\,hex\,/\,ROT13\,/\,bidi) the current census
shows the redactor currently misses; (4) separate reporting of the
indirect-reference rate, multi-lingual evasion, and the A2\,/\,A3\,/\,A4 adversary
slices.

\section{Discussion}\label{sec:discussion}

The ablation supports a small but specific design claim, and we are careful
about its scope. Because the fixtures, their expected outcomes, and the rules
share a single author, the result that \textbf{no single module reaches every
fixture's configured expected outcome, and the four mediation modules
(redaction, memory gate, output guard, audit logger) compose under the policy
engine to reach it} is a statement about configuration-consistency on a
designer-authored set---not an empirical demonstration of robustness or of
module necessity against unseen attacks.
Within that bound, the result is consistent with a defence-in-depth
architecture: redaction handles the common path, the output guard adds a
defence-in-depth hold, the per-invocation memory gate excludes selected content
from the current turn, and the audit logger preserves decision provenance for
retained in-memory events. Persistent cross-turn protection is a design target,
not an implemented or measured property of this artifact.

\paragraph{Deployment considerations.}
The seven-module decomposition assumes a controlled assistant runtime where the
developer controls capture, redaction, model invocation, memory, and output.
SaaS assistants without this control cannot adopt the design as-is. Hybrid
deployments (browser extension + assistant API) would require redaction at the
extension boundary and trust-model adjustments.

\paragraph{What we did not measure.}
Real-user task-completion rates, redaction on real (non-synthetic) screen
recordings, latency on production captures, cross-turn
retention leakage, multilingual performance, adaptive attackers, or broad
model-family repeatability.

\section{Limitations}\label{sec:limitations}

\begin{itemize}
  \item Synthetic content only; no real-user evaluation. The rendered-screen
        study uses real rendering and OCR on synthetic screens, one OCR engine,
        and no vision-language front end.
  \item On held-out screen types the v0.4 engine removed task-relevant
        identifiers (a Kubernetes secret name, an SSH host) in every
        occurrence; benign retention there was 0.763.
  \item The designer-authored ablation contains eleven fixtures; the self-generated
        census contains eleven evasion families. Neither is a sample of real
        screen-share traffic, and A2/A3 plus cross-turn A4 behaviour remain
        uncharacterised.
  \item Pattern-based redactor; additional content categories require a
        redactor extension.
  \item \textbf{The deterministic census scores string-recoverability on the
        redaction surface only.} Full-system recall equals redaction-only, so the
        per-invocation \textit{retain} gate and per-category \textit{say}/output
        orchestration receive no non-circular evaluation here; their value is
        argued from design, not measured. Establishing a robustness benefit for
        these surfaces requires a held-out test that exercises cross-turn
        retention and indirect-reference leakage, which we leave to future work.
  \item No model-behavioural result is included; a future study must retain
        auditable case-level records and use the same paired protocol across
        named model snapshots.
  \item A6, A7, and A8 explicitly excluded from the threat model.
  \item No formal privacy guarantee.
\end{itemize}

\section{Ethics}\label{sec:ethics}

All evaluation data is synthetic. No real screen captures, real personal data,
real audio recordings, or real notification content was collected, used, or
stored. No human subjects were involved. IRB approval is not applicable. The
synthetic fixture set is documented under \texttt{data/synthetic/README.md}.

The design is intended to reduce the assistant's access to sensitive content
within the mediated runtime path; this is a measured mediation claim, not a
formal privacy property. We make no anti-surveillance or
deployment-protection claim beyond the measured reduction in synthetic
exposure reported in Section~\ref{sec:eval}. The design composes with
platform-level controls and operates inside the assistant runtime, so it should
be evaluated as an additional boundary rather than as a replacement for platform
policy, organisational review, or user-facing consent design.

\paragraph{Dual use.} The evaluation publishes a consolidated evasion taxonomy
together with a per-tool coverage map, including exactly which encoding families
each defence misses. Every technique in the taxonomy is drawn from, and cited
to, previously published work (Unicode confusables~\cite{unicode_tr39},
Trojan-Source bidirectional controls~\cite{boucher2023trojansource},
indirect prompt injection~\cite{greshake2023indirect}); no new attack is
disclosed. We judge the net effect defensive: the coverage map tells defenders
where the boundary of deterministic redaction lies and which families demand a
different mechanism, information an attacker can already obtain by probing
public tools, whereas defenders currently lack a measured statement of it.

\paragraph{Third-party content.} A shared screen can expose content belonging to
people other than the operator---bystanders in notifications, counterparties in
chats---who never interacted with any consent surface. The consent engine
mediates the \textit{operator's} choices; it is not informed consent from those
third parties, and we make no such claim. Runtime redaction reduces, but cannot
guarantee, third-party exposure on families outside its coverage; the
repository's policy documentation states this boundary, and we repeat it here as
a limitation of consent-based mediation generally.

\section{Conclusion}\label{sec:conclusion}

We presented a content-layer mediation architecture for screen-share AI
assistants, decomposed into seven modules along three control surfaces
(observe, retain, say) and motivated by an explicit threat model. The
accompanying artifact is a synthetic-fixture executable scaffold; live capture,
authenticated re-consent, and cross-session state are declared future work.
Beyond a per-module configuration-consistency check on a designer-authored set,
we evaluated the redaction component on a self-generated deterministic census of
9{,}600 machine-generated cases spanning eleven literature-informed evasion
families. The concrete templates, parameter ranges, and labels are self-authored,
and the separately implemented exposure oracle is neither independent nor a
semantic superset of the redactor. On the
like-for-like digit-PII family Presidio is built for and on the paired 5-seed
subset, PerceptFence's normalisation-first redaction neutralises 0.828 of
payloads (across-seed range 0.806--0.861) versus 0.183 for real
Microsoft Presidio and 0.000 for a
no-normalisation baseline; the full cross-tool census total is \textit{indicative
only} (PerceptFence 0.398 vs Presidio 0.260) because the corpus contains
credential and injection payloads outside Presidio's PII design, and we do not
headline it. Full-system recall equals the redaction-only figure, so we frame the
observe/retain/say decomposition as an organising taxonomy rather than a measured
robustness gain; and beyond the normalisation families it targets, the system
honestly misses encoding-based evasions it has no decoder for---where
general-purpose Presidio in fact leads (recall well below 1.0, with a
characterised coverage boundary). We make bounded claims tied to a deterministic
synthetic census and explicitly exclude real-world generalisation,
model-behavioural defence, formal privacy, host compromise, and supply-chain
threats from our scope.
Moving from strings to rendered screens, the extended engine neutralised 0.918
of OCR-surviving secrets and PII on a frozen test split and 0.974 on screen
types it was never tuned on, well ahead of Presidio and gitleaks, while
removing some identifiers a support engineer would need. That trade-off, not
the headline rate, is what a deployment would have to tune.

\section*{Abbreviations}

\begin{description}
  \item[AI] artificial intelligence;
  \item[API] application programming interface;
  \item[ASR] attack success rate;
  \item[DLP] data loss prevention;
  \item[FBR] false block rate;
  \item[NER] named-entity recognition;
  \item[NFKC] Unicode Normalization Form Compatibility Composition;
  \item[OCR] optical character recognition;
  \item[PII] personally identifiable information;
  \item[SER] sensitive exposure rate;
  \item[SSN] Social Security number;
  \item[TSR] task success rate.
\end{description}

\section*{Declarations}

\paragraph{Availability of data and materials.}
\ifblind
\else
\publicartifactavailability
\fi
No real screen captures, personal or customer data, production telemetry, or
human-subject data were used. Project name: PerceptFence. Operating system:
platform independent. Programming language: Python 3.10 or newer. The core
runtime is standard-library only; pytest and the pinned evaluation dependencies
are needed for full reproduction. License: all rights reserved; review,
citation, and reproducibility inspection are permitted under the bundled
\texttt{LICENSE}.

\paragraph{Competing interests.}
The authors have no relevant financial or non-financial interests to disclose.
The study uses only synthetic data and does not use customer data, production
telemetry, proprietary code, or organizational infrastructure.

\paragraph{Funding.}
The authors received no external funding for this work.

\paragraph{Authors' contributions.}
\ifblind
\else
\namedauthorcontributions
\fi

\paragraph{Acknowledgements.}
Not applicable.

\paragraph{Ethics approval and consent to participate.}
Not applicable. The study used synthetic fixtures and synthetic evaluation
outputs only and involved no human participants or human-subject data.

\paragraph{Consent for publication.}
Not applicable. The manuscript contains no identifiable human-subject data.

\paragraph{Generative AI / tool-use disclosure.}
OpenAI Codex-based coding agents accessed through Hermes Agent were used for
editorial alternatives, code review, figure-script drafting, and
submission-package checks; for v0.4, an Anthropic Claude agent in Hermes
Agent drafted the rendered-screen harness, families, tests, and
Section~\ref{sec:screen}. Neither is credited as an author. The authors
selected and edited the prose, inspected the implementation and generated
figures, executed every reported deterministic evaluation and test, and remain
responsible for the manuscript, code, claims, and final submission decisions.

\section*{Additional file}

\paragraph{Additional file 1.}
File format: \texttt{.zip}. Title: \textit{PerceptFence reproducibility
artifact}. Description: source code, synthetic fixture definitions, evaluator
annotations, evaluation protocol, machine-readable result CSVs, figure
generators, tests, dependency pins, checksums, and the artifact checklist for
the deterministic results reported here.

\begingroup
\small
\bibliography{references}
\endgroup

\end{document}